\documentclass[review]{elsarticle}
\usepackage{lineno,hyperref}
\modulolinenumbers[5]
\journal{Journal of \LaTeX\ Templates}
\usepackage{bm}             % bold math
\usepackage{graphicx}
\usepackage{dcolumn}        % Align table columns on decimal point
\usepackage[utf8]{inputenc} % allow utf-8 input
\usepackage[T1]{fontenc}    % use 8-bit T1 fonts
\usepackage{hyperref}       % hyperlinks
\usepackage{url}            % simple URL typesetting
\usepackage{booktabs}       % professional-quality tables
\usepackage{amsfonts}       % blackboard math symbols
\usepackage{nicefrac}       % compact symbols for 1/2, etc.
\usepackage{microtype}      % microtypography
\usepackage{xcolor}
\newcommand{\red}[1]{\textcolor{black}{#1}}
\newcommand{\blue}[1]{\textcolor{black}{#1}}
\newcommand{\bra}{\left\langle}
\newcommand{\ket}{\right\rangle}
\newcommand{\bv}[1]{{\bf #1}}

\newcommand{\Dt}{\Delta t}
\newcommand{\lr}[1]{\left(#1\right)}
\newcommand{\nm}{\nonumber}

\begin{document}
\begin{frontmatter}
\title{
Crime Prediction by Data-Driven Green's Function method
% Data-Driven Green's Function method: an application of crime forecasting
}
\author[mymainaddress,mysecondaryaddress]{Mami Kajita\corref{mycorrespondingauthor}}
\cortext[mycorrespondingauthor]{Corresponding author}
\ead{kajita@csis.u-tokyo.ac.jp}
\author[mymainaddress]{Seiji Kajita}

\address[mymainaddress]{Singular Perturbations Inc, 1-5-6-5F, Kudanminami, Chiyoda-ku, Tokyo, 102-0074, Japan}
\address[mysecondaryaddress]{Center for Spatial Information Science (CSIS), the University of Tokyo, 5-1-5 Kashiwanoha, Kashiwa-shi, Chiba, 277-8568, Japan }

\begin{abstract}
We develop an algorithm that forecasts cascading events, by employing a Green's function scheme on the basis of the self-exciting point process  model.
This method is applied to open data of 10 types of crimes happened in Chicago. It shows a good prediction accuracy superior to or comparable to the standard methods which are the expectation-maximization method and prospective hotspot maps method.
We find a cascade influence of the crimes that has a long-time, logarithmic tail; this result is consistent with an earlier study on burglaries.
This  long-tail feature cannot be reproduced by the other standard methods.
In addition, a merit of the Green's function method
is the low computational cost  in the case of
high density of events and/or large amount of the training data.
%, the Green's function method is a useful component of ensemble learning.
%The present method is a powerful tool not only to predict events but also to reveal their hidden correlation.˙
\end{abstract}

\begin{keyword}
Crime forecasting; 
Green's function; 
Near repeat victimization;
self-exciting point process,
Expectation-maximization; 
Crime hotspot;
spatiotemporal forecasting
% \MSC[2010] 00-01\sep  99-00
\end{keyword}

\end{frontmatter}

% \linenumbers

%----------------------------------------------------------------------------------------------------%
%-----------------------------------------------------------------------------------%
% Introduction          													        %
%-----------------------------------------------------------------------------------%
\section{Introduction}
Prediction of crimes has been studied by two points of views
\cite{perry2013predictive}.
The first one is to predict offenders and victims,
by analyzing backgrounds of the individuals, their social class, and other epidemiological factors \cite{bonta1998, baskin2013}.
One can obtain a set of denotative explanations 
on relations between crimes and human aspects, which
 are used to screen
potential criminals or victims from a database of people recorded in polices.
This methodology is developed in the fields of
social psychology \cite{andrews1994}, sociological criminology, and psychopathology \cite{edens2009,harris2013, krakowski1986}.
The second one 
focuses on place and time of a future crime. 
Spatial information science investigates
correlations among geographical  information of criminals, victims, and environments \cite{
% cohen1979,nakaya2010,
sherman1989,anselin2000, tita2011,ohyama2018} (e.g., weather, demographics, physical environments like bars, 
parking lots, security cameras,
and even social media like 
twitter comments \cite{gerber2014}).
Mathematical models are used to extract temporal
patterns and tendencies of crime events.
For example, the approaches are composed of
pattern formation models \cite{short2008}, 
space-time  autoregressive models \cite{shoesmith2013space}, 
 network models for burglary dynamics on streets \cite{davies2013},
% model based on spatiotemporal point processes solved by Expectation-Maximization algorithms
log-Gaussian Cox Processes 
\cite{rodorigues, mohler2013, shirota2017space}, 
% \cite{flaxman2015,rodorigues,noiman, eichler,zhou}
the self-exciting point process models (SEPP)
% (Levine, 2004; Liu and Brown, 2003; Taddy, 2010; Mohler et al., 2011; Rosser and Cheng, 2016)
\cite{mohler2013,mohler, mohler2014}.
% ,and Bayesian approaches \cite{rodorigues,flaxman2015,noiman, eichler,zhou}.

In this paper, we study the SEPP model
to forecast location and time of a future crime \cite{mohler,pointprocess}.
This SEPP model captures
{\it near-repeat victimization} that is a heuristic
trend of a past crime to trigger future crimes in its vicinity \cite{johnson2007, johnson2008, short2009}.
This near-repeat victimization is based on a hypothesis
in which
criminals typically
collect information about local vulnerabilities of the targeted persons as thoroughly as possible before committing a crime.
Once the information is  collected,
they tend to repeat crimes in the vicinity of the previous one because they prefer to benefit from the already gained knowledge.

The SEPP model describes this cascading phenomena, as 
\begin{equation}
\lambda(t,\bv{x}) = \sum_{t_i<t}g(t-t_i,\bv{x}-\bv{x}_i) + \lambda_0,
\label{pointprocessmodel}
\end{equation}
where $\lambda(t,\bv{x})$ is a conditional intensity of 
a space-time point process.
This quantity indicates an expected rate of 
 number of the crime events at time $t$ and position $\bv{x}$  
  per unit time and unit area.
  % when the past data history is given.
%a crime-rate density that indicates number of crimes happened at time $t$ and position $\bv{x}$ divided by a unit time interval and area.
The $g(t-t_i,\bv{x}-\bv{x}_i)$ term corresponds to the cascading effect triggered by the past $i$-th event at $(t_i,\bv{x}_i)$, assuming that it depends only on spatiotemporal spans
between $(t,\bv{x})$ and $(t_i,\bv{x}_i)$.
The term $\lambda_0$ is a  background rate density.

%An important feature of the SEPP model is the $g$ function, 
%because it enables us
%to know a causal correlation of the events, if any, not only to predict the future %event.
%The cascading influence described by the SEPP model 
%reflects to $g(t-t_i,\bv{x}-\bv{x}_i)$  in Eq. \ref{pointprocessmodel}. 
A non-parametric approach to detemine $g$, originally applied to seismology to predict aftershocks of an earthquake, 
was achieved by a combination of
the Expectation-Maximization (EM) algorithm  \cite{marsan2008, marsan2010}.
After that, it was optimized to predict where and when crimes will occur    \cite{mohler}.
% While the non-parametric approach is usually superior to the parametric approaches \cite{adepeju2016}, a variety of non-parametric methods is still very limited.-->
While non-parametric approaches of 
spatiotemporal crime forecasting
% based on the point process models 
have been proposed
\cite{noiman,zhou,flaxman2015,eichler, mohler2017, flaxman2018},
a variety of non-parametric methods is still very limited.

In order to improve the crime-prediction tool box so as to adapt to a larger variety of situations,
we present an alternative non-parametric algorithm to determine $g$ 
inspired by an idea from material science.
 A Green's function technique is commonly used to correlate many physical properties  to external perturbations as a response of a physical system 
 \red{
 \cite{stanley, kajita2016, cai}
 }.
Given this feature of the Green's function,
the method is expected to extract the cascading influence of  crime events in the well-defined mathematical way.
 Since the Green's function is typically very difficult to
 derive \cite{kajita2016},
Cai et al. developed a scheme to reproduce it by output data of the molecular dynamics simulations by means of fluctuation dissipation theorem  \cite{cai}.
Similarly,  here we introduce the concept of the Green's function
to a data-driven approach combined with the SEPP model.

% This paper presents a non-parametric algorithm of the crime prediction, of which idea comes from materials science.
%  A numerical technique is available to determine an unknown response function from a dataset of molecular dynamics simulations.
% For example, some researchers are interested in dynamical properties of surface atoms which are connected to an infinitely large number of solid atoms.
% With an aim to reduce the computational costs, the degrees of freedom of the solid atoms can be projected on the surface atoms by using a Green's function \cite{kajita2016}.
% Since the Green's function is typically very difficult to derive,
% Cai et al. developed a scheme to reproduce a Green's function from output of the molecular dynamics simulations with a finite-size system, by means of fluctuation dissipation theorem \cite{cai}.
% Similarly, for the crime prediction, here we introduce the concept of the Green's function generated by a dataset and propose a data-driven approach combined with the SEPP model.

%----------------------------------------------------------------------------------------------------%
%-----------------------------------------------------------------------------------%
% Theory                													        %
%-----------------------------------------------------------------------------------%
\section{Theory}\label{sec:theory}

In order to design the $g(t,\bv{x})$ term by a Green's function scheme,
let us consider a density field of the crime events as
\begin{eqnarray}
\hat{\rho}(t,\bv{x}) = \sum_{i}\delta(t-t_i)\delta(\bv{x}-\bv{x}_i),
\nonumber
% \label{basic_rho}
\end{eqnarray}
where $\delta$ is the Dirac delta function.
By  using this $\hat{\rho}(t,\bv{x})$,
Eq. (\ref{pointprocessmodel}) becomes
\begin{eqnarray}
\lambda(t,\bv{x})
%\sum_{t_i<t} g(t-t_i,\bv{x}-\bv{x}_i)
 = \int_0^t dt' \int d\bv{x}' g\lr{t-t',\bv{x}-\bv{x}'} \hat{\rho}(t',\bv{x}'),
\label{point_process2}
\end{eqnarray}
where the background $\lambda_0 = 0$ because we focus on modeling the cascading influence.
Instead of the field density $\hat \rho$ which
 is composed of Dirac delta functions,
 here we use a calculable form of a  density defined as
 $\rho(t,\bv{x}) = \sum_{i\in D} \delta_{\bv{x},\bv{x}_i} \delta_{t, t_i}/ (\Delta  S \Delta t) $,
 where a subset $D$ consists of the events in the rectangular $\bv{x}$ cell
  and $|t-t_i|<\Delta t$;
$\Dt$ and $\Delta S$ are a time interval and discretized area, respectively.
By replacing $\hat{\rho}$ by the discretized density $\rho$, Eq. (\ref{point_process2}) is approximated as
\begin{eqnarray}
\lambda(t,\bv{x})
 \sim \int_0^t dt' \int d\bv{x}' g\lr{t-t',\bv{x}-\bv{x}'} \rho(t',\bv{x}').
\label{point_process3}
\end{eqnarray}

Because the conditional intensity $\lambda(t,\bv{x})$ is used to expect a 
crime density at future $t$,
an ideal $\lambda(t,\bv{x})$ should be proportional to 
$\rho(t,\bv{x})$ at the same $t$.
Moreover, the key insight is that
 the right-hand side of Eq.~(\ref{point_process3}) can be seen
as a special solution of a partial differential equation,
when we assume the $g$ as a Green's function
under an external-force field $\rho$ \cite{stanley}.
Therefore, Eq. (\ref{point_process3}) suggests that
$\rho(t,\bv{x})$ is formulated in the framework of a solution of
a differential equation as
\begin{eqnarray}
  \rho(t,\bv{x}) =
  \gamma \int_0^t dt' \int d\bv{x}' g\lr{t-t',\bv{x}-\bv{x}'}   \rho(t',\bv{x}')
  + \Delta t \int d\bv{x}' g(t,\bv{x}-\bv{x}')  \rho(t=0,\bv{x}'),
\label{ddgf}
\end{eqnarray}
where $\gamma$ is a coefficient of the feedback of the historical crime to the present; this parameter will be decided later.
The second term in the right-hand side of Eq.~(\ref{ddgf}) represents the general solution determined by
the initial condition at $t=0$
and
the boundary condition $\rho(t, |\bv{x}| \to \infty) = 0$.
The coefficient $\Delta t$ is introduced so as to make the dimension of $g$ consistent to the definition in
Eq.~(\ref{pointprocessmodel}).

%Equation.~(\ref{ddgf}) at $t=0$ becomes
%\begin{eqnarray}
% \rho(t=0,\bv{x}) =  \Delta t \int d\bv{x}' g(t=0,\bv{x}-\bv{x}')  \rho(t=0,\bv{x}'),
%\end{eqnarray}
%and then we can derive
%$g(t=0, \bv{x}) = %\delta(\bv{x})/\Delta t $, and then $g(t=0, \bv{k}) = 1/\Delta t$.

\red{
Throughout this paper,
following notations of Fourier and Laplace
transformations for an arbitrary 
function $Y$ are used unless otherwise noted.
\begin{eqnarray}
Y(\bv{k}) &=& \int \exp(\textrm{i} \bv{x} \cdot \bv{k}) Y(\bv{x}) d\bv{x}, \nm \\
Y(z) &=& \int_{0}^{\infty}
 \exp(- z t) Y(t) dt, \nm
\end{eqnarray}
where $\bv{k}$ and $z$ are two-dimensional wave-number vector and 
complex coordinate, respectively.
The Fourier and Laplace transforms 
make
convolution integrals in Eq. \ref{ddgf} be simple product forms as,
\begin{eqnarray}
\int \exp(\textrm{i} \bv{x} \cdot \bv{k}) \int 
Y’(\bv{x} - \bv{x'}) Y(\bv{x'}) d\bv{x'}d\bv{x} &=&
Y'(\bv{k}) Y(\bv{k}), \nm \\
\int_{0}^{\infty} \int_{0}^{t}\exp(- z t) Y'(t-t') Y(t') dt'  dt &=& Y'(z) Y(z). \nm
\end{eqnarray}
}
By using above the formula, 
Eq.~(\ref{ddgf}) becomes
\begin{eqnarray}
  \rho(z,\bv{k}) =   \gamma g\lr{z,\bv{k}}   \rho(z,\bv{k})
  +  \Delta t g(z,\bv{k})  \rho(t=0,\bv{k}),
\label{ddgf_kz}
\end{eqnarray}
Equation (\ref{ddgf_kz}) can be solved as
\begin{eqnarray}
  \rho(z,\bv{k}) = \Phi(z,\bv{k})   \rho(t=0,\bv{k}),
\label{rho_zk}
\end{eqnarray}
where
\begin{equation}
\Phi(z,\bv{k}) = \frac{ \Delta t g(z,\bv{k})}{1 - \gamma   g(z,\bv{k})}.
\label{phi_g}
\end{equation}
The Laplace inverse transform of Eq.~(\ref{rho_zk}) leads to
\begin{eqnarray}
 \Phi(t,\bv{k}) =\frac{  \rho(t,\bv{k}) }{  \rho(t=0,\bv{k})}.
 \nonumber
\end{eqnarray}
The function $\Phi(t,\bv{k})$
is a time development operator for the density of crime events.
\red{
While $\Phi(t,\bv{k})$
has been derived
 on the basis of a deterministic equation assumed in Eq. \ref{ddgf},
real crime events occur as a stochastic process. 
To interpolate the stochastic feature, we determine $\Phi(t,\bv{k})$ by means of the statistical average
of the whole dataset with respect to
pairs of the densities with the time difference $t$ as
}
\begin{eqnarray}
 \Phi(t,\bv{k}) =\bra\frac{  \rho(t+t_0,\bv{k}) }{  \rho(t_0,\bv{k})}\ket_{t_0},
 \label{phi}
\end{eqnarray}
where $t_0$  is the initial time for each sample in the statistical average.

Lastly, we  determine the parameter $\gamma$ so as to
 make the conditional intensity $\lambda(t,\bv{x})$
numerically equivalent to the density $ \rho(t,\bv{x})$ in a stationary state.
Suppose that crimes uniformly
occur
every unit time. The crime density at this stationary state is
\begin{eqnarray}
 \rho_{st}(t,\bv{x}) = \frac{1}{\Delta t \Delta S} 
 \nonumber
\end{eqnarray}
Through Eqs. (\ref{phi_g}) and (\ref{phi}),
one can yield
\begin{eqnarray}
 g_{st}(t,\bv{x}) = \frac{\delta(\bv{x})}{\Delta t} 
 \exp(-\gamma \frac{t}{\Delta t}).
 \nonumber
\end{eqnarray}
 Thus, Eq. (\ref{pointprocessmodel}) with $\lambda_0 = 0$ becomes
\begin{eqnarray}
\lambda_{st}(t,\bv{x})
%= \sum_{t_i<t} g_{st}(t-t_i, \bv{x}-\bv{x}_i)
= \sum_{n=1}^{\infty} g_{st}(n \Dt, 0)
&=& 
\lim_{|\bv{x}|\to 0}
\frac{\delta(\bv{x})}{\Delta t} \times \frac{1}{\exp(\gamma) -1}
\nonumber \\
&\sim& \frac{1}{\Delta t \Delta S} \times \frac{1}{\exp(\gamma) -1},
\nonumber
\end{eqnarray}
where the last approximation uses the descritization of the delta function.
As assumed that $\lambda_{st} = \rho_{st}$, 
$\gamma = \log 2$ is obtained.

In short,
$\Phi(z,\bv{k})$ is calculated through the Laplace transform of
$\Phi(t,\bv{k})$ by Eq.~(\ref{phi}).
According to Eq.~(\ref{phi_g}),
$g(z,\bv{k})$ is obtained by
\begin{equation}
g(z,\bv{k})
= \frac{\Phi(z,\bv{k})}{\Delta t + \Phi(z,\bv{k}) \log 2 }.
\label{phi_g_final}
\end{equation}
The Fourier and Laplace inverse transforms of Eq.~(\ref{phi_g_final}) give
$g(t,\bv{x})$, that
is used in Eq. (\ref{pointprocessmodel}) to predict future crimes.
In the following, we call this method 
data-driven Green's function (DDGF) method.
\red{
The DDGF method derives $g$ as a result of a {\it solution} of a partial differential equation.
This feature does not need any iteration steps for maximizing or minimizing likelihood or cost functions
as in EM method and other machine learning techniques. 
}

%----------------------------------------------------------------------%
% Calculation data
%----------------------------------------------------------------------%

\section{Computational details}\label{sec:3}

A direct calculations of Eq.~(\ref{phi}) may arise
a numerical instability, because the denominator $\rho(t_0,\bv{k})$
happens to be a very small value.
This small intensity of $\rho(t_0,\bv{k})$ results from
interferences of the Fourier components
due to simultaneous events at time $t_0$.
However, recalling that the events do not happen at exactly the same time in reality,
one can conclude that this interference is a mathematical artifact
due to assigning the multiple events happened at $|t_0-t_i|<\Delta t$
to the same $t_0$ mesh.
Based on this argument,
the instability can be fixed by a decomposition of the density field into individual crimes as
\begin{eqnarray}
\Phi(t,\bv{k})
= \bra \sum_j^{N_{t_0}} \frac{ \rho(t+t_0,\bv{k}) }{\rho_j(t_0,\bv{k}) } \ket_{t_0},
 \label{phi2}
\end{eqnarray}
 where $\rho_j(t_0, \bv{k})$ is the individual $j$-th event that satisfies
$\rho(t_0, \bv{k})=\sum_j^{N_{t_0}}\rho_j(t_0, \bv{k})$, and
$N_{t_0}$ is the number of crimes at $t_0$.

The actual steps for the DDGF algorithm are written in the followings.
\begin{description}
\item[Step 1] The discretized density $\rho(t,\bv{x})$ is generated from a dataset
\red{ 
by 
 $\rho(t,\bv{x}) = \sum_{i\in D} \delta_{\bv{x},\bv{x}_i} \delta_{t, t_i}/ (\Delta  S \Delta t) $,
 where $i$ runs in a subset $D$ that consists of the events happened in a cell at $\bv{x}$ and $|t-t_i|<\Delta t$.
 }
Then, the discrete Fourier transform is performed to obtain $\rho(t,\bv{k})$.
\item[Step 2]
$\Phi(z,\bv{k})$ is calculated by the Laplace transform of
the numerical result of 
Eq.~(\ref{phi2}).
\item[Step 3]
The Cartesian coordinates $(k_x, k_y)$  are converted
into the polar coordinates $(k_r,k_\theta)$.
Then, $\Phi(z,k_r,k_\theta)$ is averaged over the angle $k_\theta$ as: $\Phi(z,k_r) = \bra \Phi(z,k_r,k_\theta)\ket_{k_\theta}$.
\item[Step 4]
The  $g(t,r)$ is calculated by
 the Laplace inverse transform \red{\cite{laplace}} and Hankel transform
 of   Eq.~(\ref{phi_g_final}).
 %%%----- 晴司；ラプラス逆変換アルゴのレファレンス。
%
\item[Step 5] The density of the predicted crimes is obtained by $\lambda(t,\bv{x}) = \sum_{t_i<t} g(t-t_i,|\bv{x}-\bv{x}_i| )$.
\end{description}

The performance of the DDGF method is evaluated by comparisons with two representative methods which determine $g(t,r)$.
One is a simple type of the EM algorithm developed in Ref. \cite{marsan2008}.
\red{
We used the initial guess of $g(t)$ as a  exponential decaying function, and 
50 iterations 
composed of the expectation and maximization steps
are executed to converge it.
}

%計算コスト
\red{
It is worthwhile to discuss computational costs between the non-parameteric methods.
Memory allocation of the DDGF method amounts to be $O(N_m)$ where $N_m$ is total number of spatial-temporal meshes, while that of the EM method depends on $O(N^2)$ where total number of crime events is $N$.
Calculation times of the DDGF and EM are $O(N_{m}^2)$ and $O(N^2)$, which comes from bottlenecks of the Laplace/Fourier transformations and calculation of transition rates between two events, respectively.
Both $N_m$ and $N$ depend on $O(S \times T)$, where $S$ and $T$ are area and length history in the training data, respectively.
Therefore, 
the DDGF method requires the low memory usage 
in comparison to the EM method, especially in the case of high density of events and/or large amount of the training dataset.
}

Another method to determine $g$ is a parametric one which is the prospective hotspot maps (PHM) method  \cite{bowers}, 
%which assumes an inverse function of the elapsed time and distance from an event.
% The $g(t,r)$ in the PHM method is
where
\begin{equation}
g(t,r) = \frac{1}{(1+t / 7 \textrm{days}) (1+2r /  \Delta x  )},
\nonumber
\end{equation}
where $\Delta x$ is the length of the spatial mesh; namely, $\Delta S = (\Delta x)^2$.
This method has cutoff parameters 
as $g(t>t_{cut},r)=0$ and $g(t, r>r_{cut})=0$.
The original method sets
$t_{cut} = 60$ days and $r_{cut} = 0.4$ km, which are 
optimized {\it a priori} so that the PHM method predicts burglaries data of small $\Delta x$ accurately \cite{bowers}.
The length of the space and time meshes are $\Delta x=0.25$ km and  $\Delta t=1$ day. 
The total number of the spatial cells is $A$.
% ;namely the total are is equal to $A \Delta S$, where $\Delta S = \Delta x^2$.

%------------------------------------------------------------------------------------%
\section{Dataset}\label{sec:dataset}
\blue{
We choose the top-ten crime types 
that frequently happened in the open data in
Chicago, which are
downloaded via URL \cite{chicago_opendata}.
 We set conditions of the dataset  
 by using "Filter Tab" in the web site, as year duration is between 2012-2010, 
 and the ten primary types are
theft, battery, criminal damage, narcotics, other offense, assault, burglary,
motor-vehicle theft, deceptive practice and robbery.
The data contains information of date, longitude, latitude, and primary type of the crimes.
}
We set a center point
in the south area of Chicago at 41.765 latitude and -87.665 longitude.
Then, the crimes that happened 
within 5 km from the center point
from 5th May 2010 to 15th September 2011
are selected to create a dataset we use.

A calculation of the crime prediction
uses successive 400-days data  (number of the crime events ranges 1600-8200)
 chosen from the dataset.
  \red{
Then,  one-day-ahead prediction is performed.
Concretely,
in the first sample, the successive 400-days data
from 5th May 2010 to 8th June 2011 are used as the training dataset
to generate $\lambda$ on 9th June 2011.
In the second sample, 
the training 400-days data 
is shifted by 2 days ranging from 7th May 2010 to 10th June 2011
 for $\lambda$  on 11th June 2011.
We average 50 samples obtained by this procedure to evaluate accuracies of crime prediction methods.
}
 In the cases of theft, battery, and narcotics, 
 we use successive 200-days data (number of the crime events $\sim$5000-7800) because the  400-days data are too large 
 especially for the EM method 
 \red{
 owing to the large $O(N^2$) memory usage.
 }
%  The time period ranges from 21th November 2010 to 15th September 2011

\section{Results}\label{sec:rd}

\begin{figure}[]
\centering
\includegraphics[width=1.0\linewidth]{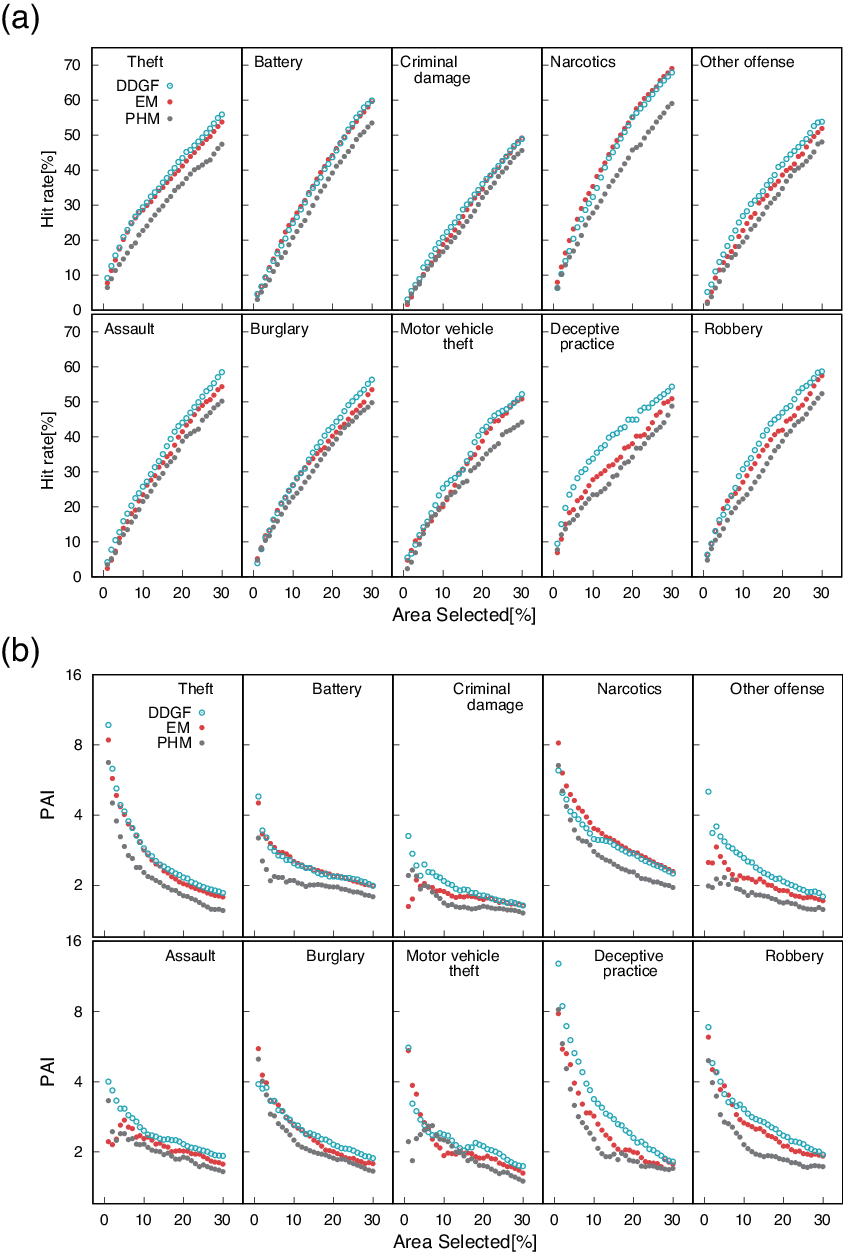}
\caption{
   (a) Hit rate and (b) PAI of the one-day-ahead crime predictions as a function of area selected.
   The results of the PHM methods are derived without cutoff radius $r_{cut}$.
The plots are obtained by averaging 50 calculations in shifting 
the prediction date once every other day,
in order to increase the statistical reliability.
  }
  \label{fig1}
\end{figure}

\begin{figure}[]
\centering
\includegraphics[width=1.0\linewidth]{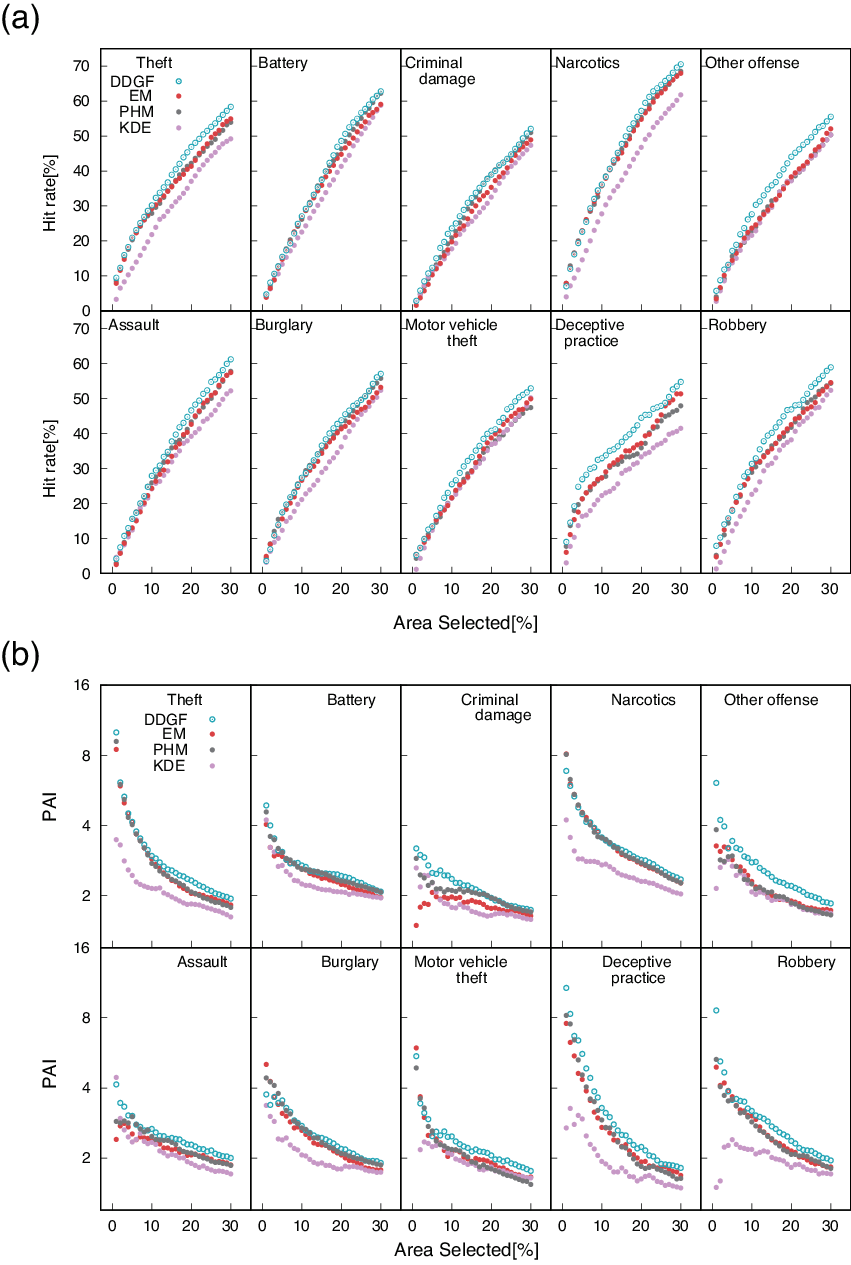}
\caption{
   (a) Hit rate and (b) PAI of the one-day-ahead crime predictions as a function of area selected
      with the cutoff radius $r_{cut} = 0.4$ km.
The plots are obtained by averaging 50 calculations in shifting 
the prediction date once every other day,
in order to increase the statistical reliability.
  }
  \label{fig2}
\end{figure}

We use two types of metrics: hit rate and predictive accuracy index (PAI).
 The cells with the higher crime rate 
$\lambda(t, \bv{x})$
are selected up to a certain percentage $a/A$ of the total area,
where $a$ is number of the selected cells.
Then, the locations of the selected area and  real crimes
are checked if they are coincident.
The hit rate is 
the correctly-predicted number of crimes divided by the total number of crimes as a function of $a/A$ which indicates
percentage of the area selected \cite{mohler,bowers}.
The PAI is defined as
the hit rate divided by $a/A$ \cite{chainey}.

\begin{table}
\caption{
Averaged hit rate and PAI 
of the one-day-ahead predictions in (upper table) Fig.\ref{fig1} and (bottom table) Fig.\ref{fig1}.
The bold number corresponds to the top accuracy between the three methods for each crime type.}
%%-----
\begin{center}
\begin{tabular}{@{}p{9em}p{3em}p{3em}p{3em}p{3em}p{3em}p{3em}@{}}
\toprule[1.0pt]
 & \multicolumn{6}{c}{without $r_{cut}$} \\
 & \multicolumn{3}{c}{Hit rate [\%]} & \multicolumn{3}{c}{PAI} \\ \cmidrule(r){2-4}\cmidrule(r){5-7}
 & DDGF & EM & PHM & DDGF & EM & PHM\\ \midrule
Theft  &\bf 36.1& 34.6 &  29.3& \bf 3.00 & 2.85 & 2.33 \\
Battery & 34.6 &\bf  35.0 & 30.1& 2.47 &\bf  2.49 & 2.06\\
Criminal damage & \bf28.3 & 27.1& 24.9 & \bf2.01 & 1.80 & 1.72\\
Narcotics & 42.9 & \bf44.6 & 36.1 & 3.18&\bf 3.48& 2.79\\
Other offense & \bf33.5 & 30.5 & 27.3 & \bf2.47 & 2.07 &1.80 \\
Assault & \bf 34.7 & 32.1, & 29.3 & \bf2.45 & 2.14 & 2.03\\
Burglary & \bf34.3 & 32.7 & 30.2 & \bf 2.52 & 2.52 & 2.28\\
Motor vehicle theft & \bf31.9 & 30.3 & 26.7 & \bf2.32 & 2.22 & 1.95\\
Deceptive practice & \bf38.5 &  32.5 & 29.1 & \bf 3.57 &  2.77 &  2.49\\
Robbery & \bf 37.8 & 35.1 & 30.7 &\bf 2.91 & 2.72 &  2.23\\
\bottomrule[1.25pt]
\end{tabular}
\end{center}
%%-----
\begin{center}
\begin{tabular}{@{}p{9em}p{3em}p{3em}p{3em}p{3em}p{3em}p{3em}p{3em}p{3em}@{}}
\toprule[1.0pt]
 & \multicolumn{8}{c}{$r_{cut} = 0.4$ km} \\
 & \multicolumn{4}{c}{Hit rate [\%]} & \multicolumn{4}{c}{PAI} \\ \cmidrule(r){2-5}\cmidrule(r){6-9}
 & DDGF & EM & PHM & \red{KDE} & DDGF & EM & PHM & \red{KDE} \\ \midrule
Theft                &\bf  37.9 & 35.3 & 35.1 &29.3 &\bf  3.11 & 2.90 & 2.91 & 2.1\\
Battery              &\bf  37.3 & 35.2 & 36.7 &32.6 &\bf  2.67 & 2.49 & 2.61 &2.31 \\
Criminal damage      &\bf  30.8 & 27.6 & 29.6 &26.2 &\bf  2.20 & 1.82 & 2.05 &1.83\\
Narcotics            &\bf  45.7 & 44.2 & 44.6 &36.7 & 3.46 &\bf  3.47 & 3.46 &2.57\\
Other offense        &\bf  34.9 & 30.5 & 30.3 &29.6 &\bf  2.67 & 2.21 & 2.18 & 2.06\\
Assault              &\bf  36.4 & 33.4 & 34.1 & 30.9 &\bf  2.53 & 2.23 & 2.33 & 2.18\\
Burglary             &\bf  34.9 & 32.9 & 34.4 &29.3 & 2.57 & 2.49 &\bf 2.59 & 2.07\\
Motor-vehicle theft  &\bf  32.5 & 29.8 & 29.0 &28.6&\bf  2.36 & 2.20 & 2.15 & 1.9\\
Deceptive practice   &\bf  37.5 & 33.2 & 32.2 &26.9&\bf  3.43 & 2.85 & 2.97 &1.96\\
Robbery              &\bf  38.1 & 34.7 & 34.7 &30.3&\bf  3.05 & 2.68 & 2.64 & 1.97\\
\bottomrule[1.25pt]
\end{tabular}
\end{center}
\label{table-oneday}
\end{table}

\begin{table}
\red{
 \caption{ Averaged hit rate and PAI of one-week-ahead prediction.
 The bold number corresponds to the top accuracy between the four methods for each crime type.}
\begin{center}
\begin{tabular}{@{}p{9em}p{3em}p{3em}p{3em}p{3em}p{3em}p{3em}p{3em}p{3em}@{}}
\toprule[1.0pt]
 & \multicolumn{8}{c}{$r_{cut} = 0.4$ km} \\
 & \multicolumn{4}{c}{Hit rate [\%]} & \multicolumn{4}{c}{PAI} \\ \cmidrule(r){2-5}\cmidrule(r){6-9}
 & DDGF & EM & PHM & \red{KDE} & DDGF & EM & PHM & \red{KDE} \\ \midrule
Theft                &\bf 37.9 & 34.8 & 34.8 & 28.9                 &\bf 3.06 & 2.77 & 2.81 & 2.04 \\
Battery              &\bf 37.3 & 34.0 & 36.4 & 32.8                 &\bf 2.65 & 2.4 & 2.59 & 2.3 \\
Criminal damage      &\bf 30.8 & 26.4 & 28.7 & 26.2                 &\bf 2.21 & 1.81 & 2.01 & 1.83 \\
Narcotics            &\bf 45.3 & 43.5 & 43.5 & 36.6                 &\bf 3.43 & 3.41 & 3.36 & 2.55 \\
Other offense        &\bf 35.5 & 30.7 & 30.8 & 30.2                 &\bf 2.7 & 2.19 & 2.19 & 2.13 \\
Assault              &\bf 35.9 & 31.0 & 33.5 & 31.0                 &\bf 2.53 & 2.1 & 2.28 & 2.2 \\
Burglary             &\bf 34.1 & 31.0 & 33.3 & 29.1                 &\bf 2.53 & 2.34 & 2.49 & 2.05 \\
Motor-vehicle theft  &\bf 32.6 & 29.5 & 29.5 & 28.2                 &\bf 2.39 & 2.19 & 2.19 & 1.88 \\
Deceptive practice   &\bf 38.8 & 32.2 & 31.4 & 26.6                 &\bf 3.41 & 2.68 & 2.72 & 1.94 \\
Robbery              &\bf 39.1 & 35.5 & 35.7 & 30.9                 &\bf 3.17 & 2.79 & 2.75 & 2.05 \\
\bottomrule[1.25pt]
\end{tabular}
\end{center}
\label{table2}
} % \red
\end{table}

Here we compare the performaces of the DDGF, EM, and PHM methods in two manners with respect to the cutoff length $r_{cut}$, where  $g(t, r>r_{cut})=0$.
Figure \ref{table-oneday} shows
the hit rate and PAI without the cutoff parameter.
Then, we introduce  $r_{cut}=0.4$ km
to these methods; the results are shown in Fig. \ref{fig2}.
Note that the original PHM method uses $r_{cut}=0.4$ km \cite{bowers}, which 
was decided so as to be equivalent to 
 the effective length of the near-repeat victimization 
 \cite{johnson2004a, johnson2004b,bowers2005}. 
Compared to this effective length of the near-repeat victimization,
$\Delta x=0.25$ km is likely to be too large, 
in spite of the fact that this mesh is the finest in the available open data in Chicago.
Owing to this large mesh,
the DDGF and EM methods may give 
an erroneous broadening of $g(t,r)$ in $r$ direction.
The introduction of $r_{cut}$, therefore, 
fixes this numerical artifacts to
increase the scores of the DDGF and EM methods, not only of the PHM method.
Indeed,
Adepeju et al. \cite{adepeju2016} studied the same open data, and they used 
the cutoff $0.3$ km for their calculations of the EM method.
Table \ref{table-oneday} shows averages of these metrics over 
whole range of the area-selected rate $a/A = 1, 2, ..., 30\%$ in each of the crime types.
The results shows that  $r_{cut}$ parameter makes the accuracies of the three methods higher.

We also evaluate performance of the one-week-ahead prediction as shown in Table \ref{table2}.
The accuracies obtained by the DDGF method are superior to or comparable to the EM and PHM methods in both one-day and one-week ahead predictions.

\section{Discussion}\label{sec:dis}

In order to clarify qualitative differences in these methods precisely, 
$g(t,r=0)$ is calculated from the larger burglary data using 450 days (from 16th March 2010 to 8th June 2011) extending the radius to 6 km.
The total number of the burglary events is 9360.
In the inset of Fig. \ref{fig2},
the DDGF and EM methods show the peaks at 1, 5, and 9 days and those at 1, 6, and 9 days, respectively.
They give very similar profiles in the short-time region.
These results indicate that the recidivism is likely to occur after the corresponding days counted from the first event.

\begin{figure}[]
\centering
\includegraphics[width=0.7\linewidth]{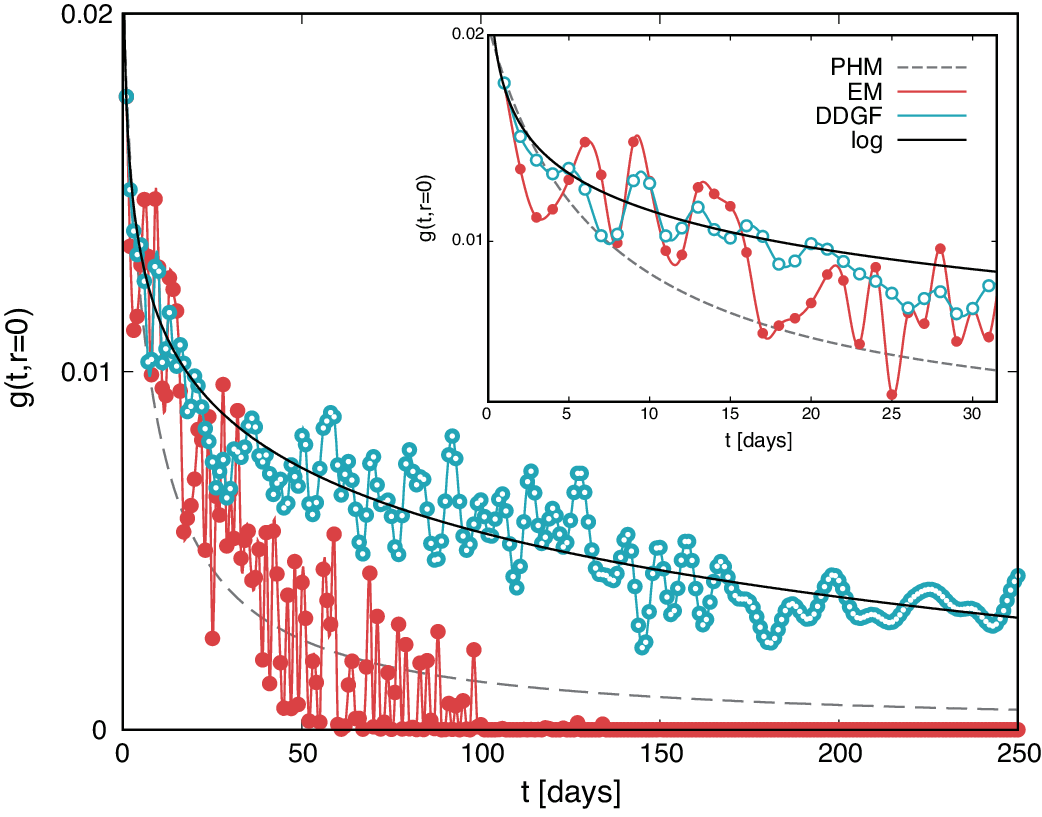}
\caption{
  Profiles of $g(t,r=0)$ functions.
  The long-time-scale feature of the DDGF method is fitted by a logarithmic function of $-a \log(b t)$, where $a$ and $b$ are fitting parameters;
  the logarithmic function (black line) is overlaid.
 The inset figure shows the short-time-scale behaviors.
  }
  \label{fig3}
\end{figure}

A distinctive feature of the DDGF method can be observed in the long-time-scale behavior of $g(t,r=0)$.
Figure  \ref{fig3} shows that $g(t,r=0)$ of the DDGF method decays very slowly in comparison to
the EM and PHM methods, and
this slow decay is well fitted by a logarithmic function (black line).
The long-time tail feature is also found in earlier studies:
some researchers collected cases of burglary that happened in the same location
in order to analyze time intervals between the events, to shed light on the near-repeat victimization \cite{anderson, burquest, ratcliffe}.
In particular, Ratcliffe et al. studied the data of burglaries  in Nottinghamshire in UK for the period of 1995-1997 \cite{ratcliffe}.
They found that the long-time-scale behavior of the observed distribution function
of the time interval is well fitted with $-\log(t)$, as observed by the DDGF method here.
Therefore, the observed logarithmic-like tail
is an evidence that the DDGF method is able to describe the causal correlation of the crime events reasonably.

\blue{
In addition to the hit rate and PAI, 
we use log-likelihoods to observe features of these models. 
The DDGF, PHM, and EM methods
indicate the log-likelihoods averaged 
over all the crime types as 0.267, 0.273, and 0.292, respectively. 
The EM method shows 
the highest log-likelihood value.
This is reasonable because the EM iteration step
iterates in such a way to maximize the log-likelihood. 
Nevertheless, from a practical view point in the crime prediction,
the hit rate and PAI are more important measures than log-likelihood.
 For example, 
 police departments want to design an efficient patrol route
 due to a limited human resource to afford patrol activities.
In this case, the log-likelihood is not suitable to be used, because
this measure refers to similarity between "overall" distributions of $\lambda$ and crime events.
On the other hand,  
the hit rate and PAI are indicators of the prediction performance
defined even at a low percentage of the selected area.
In fact, the hit rate or PAI are selected as standard measures in crime-prediction competitions and benchmark research\cite{mohler, adepeju2016,nij}, and some of the methods decide the model parameters so as to maximize these measures\cite{mohler2017}.
}

\blue{
The kernel density estimation (KDE) method are also compared.
We use the data amount for 400 days, which is the same condition as for the PHM, EM, and DDGF methods.
The bandwidth is automatically decided by a standard rule-of-thumb scheme developed by Silverman \cite{silverman}.
The results are shown in Fig. \ref{fig2}, the lower table in Table \ref{table-oneday}, and Table \ref{table2}.
The prediction scores of the DDGF, EM, and PHM methods are larger than those of the KDE.
}

\blue{
This study focused on the dynamic property described by a triggering kernel $g$ in Eq. (1)
for the cascading effect. 
The background rate $\lambda_0$, which is the second term of the right-hand side in Eq. (1),
is also an important term which describes a quasi-stationary influence of geometrical factors to the crime events.
It represents influences from  demographic factors (e.g., population and income distributions)
and crime generators/attractors (e.g., bars and clubs)\cite{block2011} in a city.
A joint use of the triggering kernel and the background rate 
is essentially useful to predict crimes;
this analysis will appear in our future work. 
}

\blue{
Here we put supplementary comments on the numerical aspects.
%With respect to predictive performance across a range of horizons,
The DDGF shows the best performance both for a day and week ahead conditions.
If considering several months ahead condition, one obtains essentially zero $\lambda$ in the cases of PHM and EM methods, because the PHM has the 60-days cut off and the EM has a negligibly small $g$, as shown in Fig. 3.
Only DDGF may show a non-zero $\lambda$ distribution due to the logarithmic long tail, 
though the $g$ is close to constant, 
%{\bf though the $\lambda$ comes to be static distribution like KDE, }
which may leads to a poor prediction performance.
With respect to the convergence sensitivity of the non-parametric model,
both of the DDGF  and EM methods are prone to produce unstable $\lambda$ in the case of small data,
due to the high degrees of freedom in the function space for $g$.
On the other hand, the parametric model is robust in terms of amount of data because of the fixed function form, though it has a poor flexibility of the representation ability of $g$.
%%
%[Adepeju 2017] adopt smoothed cross-validation bandwidth with optimal estimation of the pilot bandwidth matrix by  sum of asymptotic mean squared error [Duong and Hazelton 2005] 
%for data of spatial mesh size 250m (They call it “Prospective Kernel Density Estimate"). 
%PHM model parameter is cut off length and standardization parameter.
%KDE model parameter is model bandwidth and kernel function. 
%
%As PHM, we chose the standard parameters tuned fo crimes and adopted by previous studies \cite{bowers, mohler, adepeju, ohyama}. 
%
%We chose KDE parameters as gaussian kernel and the bandwidth for spatial scale is selected as "silverman" type using scikit-learn package. The selection of the bandwidth is widely ranged in the previous studies.
%For example, the bandwidth of 250m is selected in [Ohyama 2018] for detailed resolution data, [Chainey 2015] changed 10m-245m depending on the crime types and resolution op each data.
%If the bandwidth is optimized it may affect the accuracy, however, there are no general guidelines for selecting the size to the bandwidth in this field, 
%and the tuning of the bandwidth of KDE is out of scope for our research, we selected the most simple case of bandwidth.
}

%-----------------------------------------------------------------------------------%
% Conclusion																        %
%-----------------------------------------------------------------------------------%
\section{Conclusion}\label{sec:5}
We developed an algorithm that forecasts future crimes by learning the Green's function from past data, combining the basis of the SEPP model. 
A systematic comparison with the standard methods, the EM and PHM is performed in terms of predicting one-day and one-week-ahead predictions for the 10 types of the crime in Chicago. The results show that DDGF exhibits a good prediction accuracy superior to or comparable to the standard methods. 
Furthermore,
the DDGF method provided us with 
a characteristic long-time, logarithmic correlation of the crimes, which is consistent with the earlier study on burglaries.
%While the result of the $g(t)$ of EM and DDGF are consistent in the short-time %range,
This long-tail feature cannot be reproduced by the other methods.
Namely, the DDGF method enables us to mine the buried causal effects of crime events.

The DDGF method can be extended to multivariate in a matrix formalism. 
This extension enables us to apply
cascading network phenomena such as 
financial time series, citation networks of scientific papers, and chain-reactive increase of users of social medias (e.g., Twitter),
climatology, anomaly detection,
demand forecasting in e-commerce, and biological signal data.
The examples will appear in our future works.

%----------------------------------------------------------------------------------------------------------------------------------%
%----------------------------------------------------------------------------------------------------------------------------------%
%----------------------------------------------------------------------------------------------------------------------------------%

\section*{References}

% \bibliography{mybibfile}

\end{document}